\documentclass[prb,twocolumn,eqsecnum,amsmath,amssymb,showpacs]{revtex4}

\usepackage{bm}
\usepackage[dvips]{graphicx}

\newcommand{\be}{\begin{equation}}
\newcommand{\ee}{\end{equation}}
\newcommand{\up}{\uparrow}
\newcommand{\down}{\downarrow}

\begin{document}

\title{Staggered orbital currents in the half-filled two-leg ladder}

\author{J. O. Fj{\ae}restad}
\email{jof@physics.brown.edu}
\author{J. B. Marston}
\email{marston@physics.brown.edu}
\affiliation{Department of Physics, Brown University, Providence, 
Rhode Island 02912-1843}
                                                                               
\date{March 12, 2002}

\begin{abstract}
  Using Abelian bosonization with a careful treatment of the Klein
  factors, we show that a certain phase of the half-filled two-leg
  ladder, previously identified as having spin-Peierls order, instead
  exhibits staggered orbital currents with no dimerization.
\end{abstract}


\maketitle

\section{Introduction}

One of the most intriguing phases of strongly correlated electrons is
known variously as the ``orbital
antiferromagnet,''\cite{halperin,ners-luther,schulz89} the ``staggered
flux phase,''\cite{AM,MA,ted,elbio,ivanov,leung} or the ``$d$-density
wave.''\cite{nayak,chakravarty} It is characterized by circulating
currents which produce local magnetic moments aligned in an
antiferromagnetic (staggered) way. As a consequence, time-reversal
symmetry as well as translational and rotational symmetries are
spontaneously broken. Another phase, the ``circulating current
phase,''\cite{varma99} is somewhat similar, but does not break
translational symmetry. These phases have received considerable
attention lately, due to their possible relevance to the pseudogap
region in the phase diagram of the cuprates.\cite{varma99,chakravarty}
A recent neutron scattering experiment\cite{neutrons} on underdoped
YBa$_2$Cu$_3$O$_{6.6}$ has been interpreted\cite{chak-kee} as evidence
for these staggered orbital currents. 

\begin{figure}[h,t]
\resizebox{8cm}{!}{\includegraphics{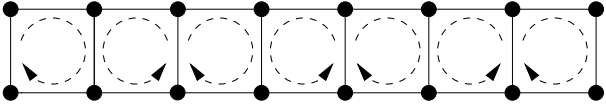}}
\caption{\label{currpatt}
Current flow in the staggered flux phase of the half-filled two-leg 
ladder. Reversing the currents gives the time-reversed state.}
\end{figure}

In this paper we focus on the half-filled two-leg ladder, which is the
simplest system that can support the staggered flux (SF) phase (see
Fig.~\ref{currpatt}). As the order parameter of this phase breaks a
discrete ($Z_2$) symmetry, the possibility of true long-range order
(LRO) of the currents is not {\em a priori} excluded at zero
temperature in this one-dimensional system, in contrast to the
situation for order parameters that break continuous symmetries, which
causes their correlations to exhibit at most quasi-LRO with power-law
decay.

For weak interactions, the ladder can be treated using bosonization
and the perturbative renormalization group (RG). For this case, the SF
phase has been found in the phase diagram for spinless electrons at
half-filling.\cite{nersesyan} Furthermore, away from half-filling,
regions with dominant tendencies toward SF ordering have been found
both for spinless\cite{ners-luther-kus} and spinful\cite{schulz96}
electrons.  Note that for general (i.e., incommensurate) fillings, true
LRO of the currents is not possible due to the absence of Umklapp
interactions (see Sec. \ref{disc} for a more detailed discussion).
The results for the doped ladder with and without spin were summarized
in Ref.~\onlinecite{orignac}, which also investigated the effects of
disorder.

Here we are concerned with the SF phase for spinful electrons in a 
weakly interacting half-filled two-leg ladder. In contrast to the
other weak-interaction studies mentioned so far, in the approach used
here the nearest-neighbor hopping parameters $t_{\perp}$ and $t$ along
the rungs and legs, respectively, can be of the same order. We
reanalyze the nature of a specific phase found in Ref.
\onlinecite{LBF98}, and demonstrate that this phase, previously
identified to be of spin-Peierls (SP) type, actually exhibits
staggered orbital currents with no dimerization, and therefore in fact
is the SF phase. In contrast to the infinite-U half-filled case, where
the constraint of no double occupancies makes the currents an
unobservable gauge artifact, the LRO currents found here are
observable. Furthermore, since all excitations are gapped, the SF
phase at half-filling is insulating.

We perform our calculations using Abelian bosonization, paying careful
attention to the Klein factors in this
formalism.\cite{revschulz98,vondelft} As a check of our treatment, we
also reproduce the identification of the CDW phase found in Ref.
\onlinecite{LBF98}. Furthermore, we show that our results are
consistent with those found for the doped ladder.\cite{orignac}

The paper is organized as follows. In
Secs.~\ref{ladder} and \ref{bosonization} we discuss the ladder model and
its continuum limit and bosonized form, closely following the approach
of Ref.~\onlinecite{LBF98}. In Sec.~\ref{sf} we define various local
order parameters, derive their bosonized expressions, and calculate
their expectation values in the SF phase.  The results are discussed
further in Sec.~\ref{disc}. Some of the technical details have been
placed in two appendices.

\section{The half-filled two-leg ladder and its continuum limit}
\label{ladder}

\subsection{Kinetic energy}

We consider a two-leg ladder where the electrons can hop
only between nearest-neighbor sites along the rungs and legs. The
kinetic energy then reads
\be
H_0 = -t\sum_{\ell m s}
c^{\dagger}_{\ell s}(m+1)c_{\ell s}(m)
 - t_{\perp}\sum_{m s}
c^{\dagger}_{1s}(m)c_{2s}(m)+\text{H.c.}
\label{KE}
\ee
The operators $c_{\ell s}(m)$ and $c^{\dagger}_{\ell s}(m)$, 
respectively, annihilate and create an electron on site $m=1,\ldots,N$
on leg $\ell=1,2$ with spin $s=\uparrow,\downarrow$, and obey
$\{c_{\ell s}(m),c^{\dagger}_{\ell's'}(m')\}
=\delta_{\ell\ell'}\delta_{mm'}\delta_{ss'}$, with all other
anticommutators vanishing. Periodic (open) boundary conditions are
used along (perpendicular to) the leg direction.  Introducing even and
odd combinations 
\be
c_{\text{e/o},s}=\frac{1}{\sqrt{2}}\left(c_{1s}\pm c_{2s}\right)
\ee
and Fourier transforming along the leg direction, the kinetic energy
becomes diagonal in momentum space, describing two uncoupled bands
with dispersion $\varepsilon_{\text{e/o}}(k)=-2t\cos ka\mp t_{\perp}$,
where $a$ is the lattice constant. Taking $t_{\perp}$ positive, the
even (odd) combination gives a bonding (antibonding) band. We consider
a half-filled system and $t_{\perp}<2t$, in which case the Fermi level
is at zero energy and crosses both bands, thus giving rise to four
Fermi points $\pm k_{F\text{e/o}}$ which satisfy
$k_{F\text{e}}+k_{F\text{o}}=\pi/a$.

We will assume weak interactions and focus on the low-energy,
long-wavelength properties of the model, so that we may linearize
$H_0$ around the Fermi points.  It will be most convenient to work
with a coordinate-space representation of the Hamiltonian. For this purpose
we decompose the band operator $c_{\lambda s}(m)$
($\lambda=\text{e},\text{o}$) into a sum of left- and right-moving slowly
varying (on the scale of the lattice constant) continuum fields,
\be
c_{\lambda s}(m)=\sqrt{a}\left[e^{-ik_{F\lambda}x}\psi_{L\lambda s}(x)+
e^{ik_{F\lambda}x}\psi_{R\lambda s}(x)\right],
\label{RLdecompose}
\ee
where $x\equiv ma$. The linearized kinetic energy can then be written
$H_0= \int dx{\cal H}_0$, where
\be
{\cal H}_0 = -iv_F\sum_{\lambda s}\left[
\psi^{\dagger}_{R\lambda s}\partial_x\psi_{R\lambda s}
-\psi^{\dagger}_{L\lambda s}\partial_x\psi_{L\lambda s}\right].
\label{KEdensity}
\ee
In this expression and throughout the paper it is understood that
products of fermionic (and bosonic) operators that may be evaluated at
the same point are to be normal-ordered. The bare Fermi velocity $v_F$
is the same for both bands and is given by
$v_F=a\sqrt{(2t)^2-t_{\perp}^2}$.

\subsection{Interactions}

The continuum description of general, but weak, finite-ranged,
spin-independent interactions, to leading order in the interaction
strengths, was carefully discussed in Refs.~\onlinecite{LBF98},~
\onlinecite{BF96}, and~\onlinecite{LBF97}.  One can restrict 
attention to terms which are both marginal (i.e., consisting of
four-fermion interactions with no spatial derivatives) and nonchiral
(i.e., containing two right-moving and two left-moving fermions). These
terms can be classified according to whether they conserve momentum or
not. The Hamiltonian density for momentum-conserving terms reads
\begin{eqnarray}
\text{$\mathcal{H}$}^{(1)}_I &=& \sum_{\lambda\mu}\Big\{
b^{\rho}_{\lambda\mu} J_{R\lambda\mu}J_{L\lambda\mu} 
- b^{\sigma}_{\lambda\mu}
\bm{J}_{R\lambda\mu}\cdot \bm{J}_{L\lambda\mu} \nonumber \\ 
 &+& f^{\rho}_{\lambda\mu} J_{R\lambda\lambda}
J_{L\mu\mu} - f^{\sigma}_{\lambda\mu} 
\bm{J}_{R\lambda\lambda}\cdot \bm{J}_{L\mu\mu}\Big\}.
\label{intdensity}
\end{eqnarray}
Here $f$ and $b$ refer to forward and backward scattering processes,
respectively, and 
\begin{eqnarray}
J_{P\lambda\mu} &=& \sum_{s}
\psi^{\dagger}_{P\lambda s}\psi_{P\mu s}, \\
\bm{J}_{P\lambda\mu} &=& \frac{1}{2}\sum_{ss'}
\psi^{\dagger}_{P\lambda s}\bm{\sigma}_{ss'}\psi_{P\mu s'},
\end{eqnarray}
where $\sigma^x$, $\sigma^y$, and $\sigma^z$ are the Pauli matrices.
The following general symmetries hold:
$b^{\nu}_{\text{eo}}=b^{\nu}_{\text{oe}}$ and
$f^{\nu}_{\text{eo}}=f^{\nu}_{\text{oe}}$, where $\nu=\rho,\sigma$.
To avoid double-counting, we set $f^{\nu}_{\lambda\lambda}= 0$.  At
half-filling the model also has particle-hole symmetry, which implies
$b^{\nu}_{\text{oo}}=b^{\nu}_{\text{ee}}$, leaving six independent
couplings of this type.

Half-filling also allows for non-momentum-conserving (i.e., Umklapp) terms.
The Hamiltonian density for these interactions reads
\be
{\cal H}_I^{(2)} = \sum_{\lambda\mu}\Big\{u^{\rho}_{\lambda\mu}
I^{\dagger}_{R\lambda\mu}I_{L\bar{\lambda}\bar{\mu}}
-u^{\sigma}_{\lambda\mu}\bm{I}^{\dagger}_{R\lambda\mu}\cdot
\bm{I}_{L\bar{\lambda}\bar{\mu}} + \text{H.c.}\Big\},
\ee
where $\bar{\text{e}}=\text{o}$ and $\bar{\text{o}}=\text{e}$. Here we have
defined
\begin{eqnarray}
I_{P\lambda\mu} &=& \sum_{ss'}\psi_{P\lambda s}
\epsilon_{ss'}\psi_{P\mu s'}, 
\label{Umcurrrho}\\
\bm{I}_{P\lambda\mu} &=& \frac{1}{2}\sum_{ss'}
\psi_{P\lambda s}(\epsilon\bm{\sigma})_{ss'}\psi_{P\mu s'},
\label{Umcurrsigma}
\end{eqnarray}
where $\epsilon=-i\sigma^y$. We may take 
$u^{\nu}_{\text{eo}}=u^{\nu}_{\text{oe}}$ since 
$I_{P\lambda\mu}=I_{P\mu\lambda}$ and
$\bm{I}_{P\lambda\mu}=-\bm{I}_{P\mu\lambda}$. The latter result
also implies $\bm{I}_{P\lambda\lambda}=0$, so that we can take 
$u^{\sigma}_{\lambda\lambda}=0$. In addition, particle-hole symmetry gives
$u^{\rho}_{\text{ee}}=u^{\rho}_{\text{oo}}$, leaving 
three independent Umklapp couplings. Thus a total of nine independent 
coupling constants must be taken into account in this model of the 
half-filled two-leg ladder. 
 
\section{Bosonization}
\label{bosonization}

In the Abelian bosonization formalism,\cite{revschulz98,vondelft,shankar} 
the fermionic field operators $\psi_{P\lambda s}$ can be expressed in
terms of dual Hermitian bosonic fields $\phi_{\lambda s}$ and
$\theta_{\lambda s}$ as\cite{diffnot}
\be
\psi_{P\lambda s}= \frac{1}{\sqrt{2\pi\epsilon}}\kappa_{\lambda s}
\exp{[i(P\phi_{\lambda s}+\theta_{\lambda s})]},
\label{bosfieldop}
\ee
where $\epsilon$ is a short-distance cutoff, and $P=R/L=\pm 1$. 
The bosonic fields satisfy the commutation relations
\begin{subequations}
\label{bosonfieldcommrel}
\begin{eqnarray}
[\phi_{\lambda s}(x),\phi_{\lambda's'}(x')] &=& [\theta_{\lambda
s}(x),\theta_{\lambda's'}(x')] = 0,\label{commphiphi}\\
\text{}[\phi_{\lambda s}(x),\theta_{\lambda's'}(x')] &=& i\pi 
\delta_{\lambda\lambda'}\delta_{ss'}\Theta(x-x'),
\label{commphitheta}
\end{eqnarray}
\end{subequations}
the latter result written for $\epsilon\to 0$. Here $\Theta(x)$ is
the Heaviside function. The long-wavelength normal-ordered fermionic
densities can be expressed in terms of the bosonic fields as
$\psi^{\dagger}_{P\lambda s}\psi_{P\lambda s} = \partial_x
(\phi_{\lambda s}+P\theta_{\lambda s})/2\pi$.

The Klein factors $\kappa_{\lambda s}$ commute with the bosonic
fields, and satisfy
\be
\{\kappa_{\lambda
s},\kappa_{\lambda's'}\}=2\delta_{\lambda\lambda'}\delta_{ss'}.
\label{anticommKlein}
\ee
Note that the Klein factors used here are Hermitian (instead of
unitary), since we follow the common procedure of neglecting the
number-changing property of the Klein factors in the thermodynamic
limit.\cite{revschulz98}

Charge and spin operators are now defined as
\begin{subequations}
\label{chargespin}
\begin{eqnarray}
\phi_{\lambda\rho} &=& \frac{1}{\sqrt{2}}(\phi_{\lambda\up}
+\phi_{\lambda\down}),\label{charge}\\
\phi_{\lambda\sigma} &=& \frac{1}{\sqrt{2}}
(\phi_{\lambda\up}-\phi_{\lambda\down}),\label{spin}
\end{eqnarray}
\end{subequations}
with similar definitions of the $\theta$ operators. We also define
\be
\phi_{r\nu}=\frac{1}{\sqrt{2}}(\phi_{\text{e}\nu}+r\phi_{\text{o}\nu}),
\label{plusminus}
\ee
where $r=\pm$ and $\nu=\rho,\sigma$. Again, similar definitions apply
to the $\theta$ operators. Both Eqs. (\ref{chargespin}) and
(\ref{plusminus}) are unitary transformations, which implies that the
commutation relations for the new sets of operators are of the same
type as those in Eq. (\ref{bosonfieldcommrel}). 

Next we consider the bosonized form of the Hamiltonian density ${\cal
H}={\cal H}_0+{\cal H}_I^{(1)}+{\cal H}_I^{(2)}$, which is most
succinctly expressed in terms of the variables $\phi_{r\nu}$ and
$\theta_{r\nu}$. The kinetic-energy density reads
\be
{\cal H}_0 = \frac{v_F}{2\pi}\sum_{r\nu}\left[(\partial_x
\phi_{r\nu})^2+(\partial_x\theta_{r\nu})^2\right].
\label{H0boson}
\ee
The momentum-conserving part of the interactions can be written
${\cal H}_I^{(1)}={\cal H}_I^{(1a)}+{\cal H}_I^{(1b)}$, where 
\be
{\cal H}_{I}^{(1a)}=\frac{1}{2\pi^2}\sum_{r\nu}A_{r\nu}
\left[(\partial_x\phi_{r\nu})^2-(\partial_x
\theta_{r\nu})^2\right].
\label{HI1a}
\ee
Here $A_{r\nu}=h_{\nu}\left[b^{\nu}_{\text{ee}}+rf^{\nu}_{\text{eo}}\right]$ 
with 
$h_{\rho}=1$, $h_{\sigma}=-1/4$. Furthermore, 
\begin{eqnarray}
\lefteqn{
{\cal H}_I^{(1b)} = -\frac{1}{(2\pi\epsilon)^2}\Bigl[2\hat{\Gamma}
b^{\sigma}_{\text{eo}}\cos 2\theta_{-\rho}\cos 2\phi_{+\sigma}} \nonumber \\ 
 &-& \cos
2\phi_{+\sigma}(2 b^{\sigma}_{\text{ee}}\cos 2\phi_{-\sigma}+2\hat{\Gamma}
 f^{\sigma}_{\text{eo}}\cos 2\theta_{-\sigma}) \nonumber \\ &+&
\cos 2\theta_{-\rho} (\hat{\Gamma}
b^+_{\text{eo}}\cos 2\phi_{-\sigma}+b^-_{\text{eo}} \cos
2\theta_{-\sigma})\Big],
\label{HI1b}
\end{eqnarray}
with $b^{\pm}_{\text{eo}}=b^{\sigma}_{\text{eo}}\pm 4
b^{\rho}_{\text{eo}}$ and $\hat{\Gamma}=\kappa_{\text{e}\up}
\kappa_{\text{e}\down}
\kappa_{\text{o}\up}\kappa_{\text{o}\down}$. Finally, the bosonized form of
the Umklapp interaction density reads
\begin{eqnarray}
\lefteqn{
{\cal H}_I^{(2)} = -\frac{2}{(2\pi\epsilon)^2} \cos 2\phi_{+\rho}\Bigl[
8\hat{\Gamma} u^{\rho}_{\text{ee}}\cos 2\theta_{-\rho} } \nonumber \\ & & 
 \hspace{-1.2cm} + 2 u^{\sigma}_{\text{eo}} \cos 2\phi_{+\sigma} +  
 u^+_{\text{eo}} \cos2\phi_{-\sigma}+\hat{\Gamma} 
u^-_{\text{eo}}\cos2\theta_{-\sigma} \Bigr],
\label{HI2}
\end{eqnarray}
with $u^{\pm}_{\text{eo}}=u^{\sigma}_{\text{eo}}\pm 4
u^{\rho}_{\text{eo}}$. 

Since the Hermitian operator $\hat{\Gamma}$ obeys $\hat{\Gamma}^2=I$, 
$\hat{\Gamma}$ has eigenvalues $\Gamma=\pm 1$. Furthermore, since
$[H,\hat{\Gamma}]=0$, $H$ and $\hat{\Gamma}$ can be simultaneously
diagonalized.

\section{The staggered flux phase}
\label{sf}

\subsection{Pinned fields}
\label{pinned}

Numerical integration of the one-loop RG equations for the couplings
shows\cite{BF96,LBF97,LBF98} that some of the couplings remain small,
while the others grow (sometimes after a sign change) and eventually
diverge. The weak-coupling RG flow must be cut off before it leaves
the regime of its perturbative validity. The ratios of the diverging
couplings at the cutoff scale are found to approach fixed constants in
the limit of asymptotically small bare couplings, with different sets
of ratios corresponding to different phases of the ladder. In the SF
phase $b^{\rho}_{\text{ee}}$ and $b^{\sigma}_{\text{ee}}$ are negligible,
while the diverging couplings are given by\cite{LBF98}
\begin{eqnarray}
 & & f^{\rho}_{\text{eo}} = -\frac{1}{4}f^{\sigma}_{\text{eo}}= -
b^{\rho}_{\text{eo}} = \frac{1}{4}b^{\sigma}_{\text{eo}} = \nonumber \\ 
 & & \frac{1}{2}u^{\sigma}_{\text{eo}}=-2u^{\rho}_{\text{eo}}= 2
u^{\rho}_{\text{ee}} \equiv g>0.
\label{ratios}
\end{eqnarray}
The resulting low-energy effective Hamiltonian can be mapped onto
an SO(8) Gross-Neveu model, whose integrability may be
exploited to extract the exact energies, degeneracies and quantum
numbers of all the low-energy excited states.\cite{LBF98} However, for
our discussion, a semiclassical reasoning will suffice.  Since the
single coupling constant $g$ flows toward large values, in the
semiclassical ground state the bosonic fields in the Hamiltonian will
be pinned to values which minimize the cosine interaction ${\cal
H}_I^{(1b)}+{\cal H}_I^{(2)}$. Note that this argument would not be
valid if the cosine interactions were to contain both the dual fields
$\phi_{-\sigma}$ and $\theta_{-\sigma}$, since then the uncertainty
principle would forbid both fields to be pinned. However,
$\phi_{-\sigma}$ disappears from the cosine interaction because
$b^{\sigma}_{\text{ee}}$ is negligible and
$b^+_{\text{eo}}=u^+_{\text{eo}}=0$. The pinned fields are then
$\phi_{+\rho}$, $\phi_{+\sigma}$, $\theta_{-\rho}$, and
$\theta_{-\sigma}$. Since all four bosonic modes $r\nu$ are pinned,
the SF phase has no gapless excitations at half-filling.

The possible solutions for the pinned fields are found by minimizing
$\langle \Gamma|H|\Gamma\rangle$, where $|\Gamma\rangle$ is the
eigenstate of $\hat{\Gamma}$ with eigenvalue $\Gamma$ (these solutions
will depend on $\Gamma$, but the physics will of course not, as will
be seen explicitly in Sec. \ref{OP}). There are infinitely many
solutions for the pinned fields that minimize $\langle
\Gamma|H|\Gamma\rangle$. However, this multitude of solutions is only
apparent; taking into account the fact that the bosonic fields are not
gauge-invariant, it can be shown that there are only two physically
distinct ground states.\cite{LBF98} The pinned-field configurations
that we will use to specify these ground states are given in Table
\ref{pinnedvalues}.

\begin{table}
\begin{ruledtabular}
\begin{tabular}{|l|r|c|c|c|c|}
Ground state & $\Gamma$ & $\langle\phi_{+\rho}\rangle$ &
$\langle\phi_{+\sigma}\rangle$ &
$\langle \theta_{-\rho}\rangle$ & $\langle\theta_{-\sigma}\rangle$ \\ \hline
SF 1   & $1$  & $0$    & $0$ & $0$     & $0$   \\
SF 2   & $1$  & $\pi$  & $0$ & $0$     & $0$   \\\hline

SF 1   & $-1$ & $\pi$  & $0$ & $\pi/2$ & $\pi/2$ \\
SF 2   & $-1$ & $0$    & $0$ & $\pi/2$ & $\pi/2$ \\\hline

CDW 1  & $1$  & $0$    & $0$ & $\pi/2$ & $0$   \\
CDW 2  & $1$  & $\pi$  & $0$ & $\pi/2$ & $0$   \\\hline

CDW 1  & $-1$ & $0$    & $0$ & $0$     & $\pi/2$ \\
CDW 2  & $-1$ & $\pi$  & $0$ & $0$     & $\pi/2$ \\ 
\end{tabular}
\end{ruledtabular}
\caption{\label{pinnedvalues}
The $\Gamma$-dependent pinned-field configurations used for
the ground states in the SF and CDW phases (we choose 
$\langle\Gamma|\kappa_{\text{e}\up}\kappa_{\text{o}\up} |\Gamma\rangle
=i$; see Appendix \ref{matelKlein}). The two configurations listed
here for a given ground state are physically equivalent, as can be
seen from Table \ref{OPresults}.}
\end{table}

\subsection{Order parameters}
\label{OP}

\begin{figure}
\vspace{0.4cm}
\resizebox{8cm}{!}{\includegraphics{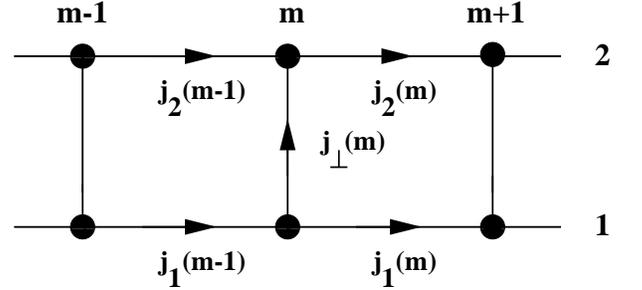}}
\caption{\label{currconservfig}
Currents as defined in Eqs. (\ref{jrung}) and (\ref{jchain}). Current
conservation is expressed by Eq. (\ref{currconserv}).}
\end{figure}

In this subsection we explicitly show that the phase characterized by
the couplings in Eq. (\ref{ratios}) is not of spin-Peierls type with a
$(\pi,\pi)$ modulation in the kinetic energy,\cite{LBF98} but instead
is the SF phase. We first define the relevant order parameters. The
fundamental definition of the current operator comes from interpreting
the Heisenberg equation of motion for the number operator
\be
n_{\ell}(m)=\sum_s c^{\dagger}_{\ell s}(m)c_{\ell s}(m)
\label{numberop}
\ee
as a discretized continuity equation. We will assume that the SF phase
is a low-energy phase of a lattice Hamiltonian whose interactions
commute with $n_{\ell}(m)$. This is, e.g., the case for density-density
and spin-exchange interactions. Then the components of the current
operator take their conventional forms (see Fig.~\ref{currconservfig})
\begin{eqnarray}
j_{\perp}(m) &=& it_{\perp} a
\sum_s\left[c^{\dagger}_{2s}(m)c_{1s}(m)-\text{H.c.}\right],
\label{jrung}\\
j_{\ell}(m) &=& it a \sum_s\left[c^{\dagger}_{\ell s}(m+1)c_{\ell
s}(m) -\text{H.c.}\right].
\label{jchain}
\end{eqnarray}
Furthermore, the local kinetic-energy operator is
\be
k_{\ell}(m)=-t\sum_s\left[c^{\dagger}_{\ell s}(m+1)c_{\ell
s}(m)+\text{H.c.}\right].
\label{localKEop}
\ee
For completeness, in our discussion we also include the number
operator $n_{\ell}(m)$ itself, since we will later check that our
calculations reproduce the results for the CDW phase found in Ref.
\onlinecite{LBF98}.

Next we outline the derivation of the bosonized expressions for these
order parameters. It will be convenient to define an auxiliary operator,
\be 
G_{\ell}(m,u,v)=\sum_s\left[c^{\dagger}_{\ell s}(m+u)c_{\ell s}(m)+v\cdot
\text{H.c.}\right].
\label{defG}
\ee
Then $j_{\ell}(m) = itaG_{\ell}(m,1,-1)$, $k_{\ell}(m) =
-tG_{\ell}(m,1,1)$, and $n_{\ell}(m) = G_{\ell}(m,0,1)$.  The
continuum version of $G_{\ell}(m,u,v)$ will contain products of type
$:\psi^{\dagger}_{P\lambda s}(x+ua)\psi_{P'\lambda' s'}(x):$,
i.e., with the argument of the field operators differing by a lattice
constant when $u=1$ (here we have temporarily included the
normal-ordering symbol explicitly). One can safely Taylor-expand
within the normal-ordering symbol to obtain $:\psi^{\dagger}_{P\lambda
s}(x)\psi_{P'\lambda' s'}(x):+ua:\partial_x \psi^{\dagger}_{P\lambda
s}(x)\psi_{P'\lambda's'}(x):$ (note that due to the normal-ordering,
all order parameters will be measured with respect to their values in
the noninteracting ground state). For now, we only keep the zeroth-order 
term in the Taylor expansion, and comment briefly on higher-order terms 
later. We find
\begin{widetext}
\vspace{-0.5cm}
\begin{eqnarray}
\lefteqn{
G_{\ell}(m,u,v) = \frac{a}{2}\sum_{P\lambda s} 
\Bigl\{\psi^{\dagger}_{P\lambda s}\psi_{P\lambda s}
\Bigl(e^{-iPk_{F\lambda}ua}+v\,e^{iPk_{F\lambda}ua}\Bigr)
 + \psi^{\dagger}_{P\lambda s}\psi_{-P\lambda
s}\;e^{-2iP k_{F\lambda}x} e^{-iPk_{F\lambda}ua}(1+v) - (-1)^{\ell}\cdot} 
\nonumber \\ & & \hspace{-1.2cm}
\Bigl[\psi^{\dagger}_{P\lambda s}\psi_{P\bar{\lambda}s}\;
e^{-iP(k_{F\lambda}-k_{F\bar{\lambda}})x} 
\Bigl(e^{-iPk_{F\lambda}ua}+v\, e^{iPk_{F\bar{\lambda}}ua}\Bigr)
 + \psi^{\dagger}_{P\lambda s}\psi_{-P\bar{\lambda}s}
e^{-iP(k_{F\lambda}+k_{F\bar{\lambda}})x}
\Bigl( e^{-iPk_{F\lambda}ua}+v\,e^{-iPk_{F\bar{\lambda}}ua}\Bigr)\Bigr]
\Bigr\}.
\end{eqnarray}
\end{widetext}
The expectation value of the normal-ordered long-wavelength density 
$\psi^{\dagger}_{P\lambda s}\psi_{P\lambda
s}$ is zero. Bosonizing $\psi^{\dagger}_{P\lambda
s}\psi_{-P\lambda s}$ produces exponentials
containing the fields $\phi_{+\rho}$, $\phi_{+\sigma}$,
$\phi_{-\rho}$ and $\phi_{-\sigma}$. Bosonizing
$\psi^{\dagger}_{P\lambda s}\psi_{P\bar{\lambda}s}$ 
produces exponentials containing the fields $\theta_{-\rho}$,
$\theta_{-\sigma}$, $\phi_{-\rho}$ and $\phi_{-\sigma}$. Since
$\phi_{-\rho}$ and $\phi_{-\sigma}$ are dual to the pinned fields
$\theta_{-\rho}$ and $\theta_{-\sigma}$, respectively, they will
fluctuate strongly due to the uncertainty principle, and the
expectation value of exponentials of these fields will vanish. We are
therefore left with a term which contains products of type
$\psi^{\dagger}_{P\lambda s}\psi_{-P\bar{\lambda}s}$. Bosonizing this
produces exponentials containing the four pinned fields, so this term
will have a nonzero expectation value.

Next consider $j_{\perp}(m)$. Its continuum expression only contains
products of type $\psi^{\dagger}_{P\lambda s}\psi_{P\bar{\lambda} s}$
and $\psi^{\dagger}_{P\lambda s}\psi_{-P\bar{\lambda}s}$. Thus only
the latter product will contribute to the expectation value of this
operator. Note that in order to calculate $j_{\perp}(m)$ no Taylor
expansion is necessary, since both fermion operators are taken at the
same value of $m$ from the outset.

Using $k_{F\text{e}}+k_{F\text{o}}=\pi/a$ and $2t\cos
k_{F\text{o}}a=t_{\perp}$ to simplify expressions, we find
\begin{eqnarray}
\hspace{-1.0cm}
\langle j_{\perp}(m)\rangle  \! &=& \! it_{\perp}a^2(-1)^m \langle
F_{-1}(x)\rangle + \text{c.c.},
\label{expjperp}\\
\hspace{-1.0cm}
\langle j_{\ell}(m)\rangle \!  &=& \! \frac{1}{2}it_{\perp}a^2(-1)^{\ell+m}
\langle F_{-1}(x)\rangle+\text{c.c.},
\label{expjell}\\
\hspace{-1.0cm}
\langle k_{\ell}(m)\rangle   \! &=& \!  
ia\sqrt{t^2-(t_{\perp}/2)^2}(-1)^{\ell+m}
\langle F_1(x)\rangle +\text{c.c.},
\label{expkell}\\
\hspace{-1.0cm}
\langle n_{\ell}(m)\rangle  \! &=& \! -a(-1)^{\ell+m}
\langle F_{1}(x)\rangle + \text{c.c.}
\label{expnell}
\end{eqnarray}
Here we have defined the operator
\be
F_p(x)=\sum_s \left[
\psi^{\dagger}_{L\text{e}s}\psi_{R\text{o}s}+p\psi^{\dagger}_{L\text{o}s}
\psi_{R\text{e}s}\right].
\ee
The expectation value of $F_p(x)$ is independent of $x$. It then follows 
from Eqs. (\ref{expjperp}) and (\ref{expjell}) that if currents exist,
they will flow around the plaquettes in a staggered pattern as shown in
Fig.~\ref{currpatt}. Current conservation is expressed by 
\be 
\langle j_{\ell}(m)\rangle =
\langle j_{\ell}(m-1)\rangle + (-1)^{\ell}\langle j_{\perp}(m)\rangle
\label{currconserv}
\ee
(see Fig.~\ref{currconservfig}).  Since expression
(\ref{expjell}) for $\langle j_{\ell}(m)\rangle$ only contains the
zeroth-order term in the Taylor series expansion of the field
operators, we conclude from Eq. (\ref{currconserv}) that the higher-order
terms do not contribute to the plaquette currents.

The bosonized expression for $F_p(x)$ can be written
\begin{eqnarray}
\lefteqn{
F_p(x) = \frac{1}{2\pi\epsilon}\sum_{\lambda s}
d_{\lambda\bar{\lambda}}^{\alpha(p)}\kappa_{\lambda
s}\kappa_{\bar{\lambda}s} }\nonumber \\ & & \hspace{-0.7cm}
\exp{\left[i\left(\phi_{+\rho}+s\phi_{+\sigma}-
d_{\lambda\bar{\lambda}}\theta_{-\rho}-s d_{\lambda\bar{\lambda}} 
\theta_{-\sigma}\right)\right]}.
\label{Fpboson}
\end{eqnarray}
Here we have defined $d_{\text{eo}}=-d_{\text{oe}}=1$, $s=\up\down=\pm
1$, and
\be
\alpha(p)=\left\{\begin{array}{lll}
1, & & p=-1, \\
2, & & p=1.
\end{array}\right.
\ee
Let $|n;\Gamma\rangle \equiv |n(\Gamma)\rangle \otimes |\Gamma\rangle$
be a simultaneous eigenstate of $H$ and $\hat{\Gamma}$. The eigenstate
$|n(\Gamma)\rangle$ lives in the Hilbert space where the bosonic
operators act, while $|\Gamma\rangle$ was introduced in
Sec.~\ref{pinned}. We now consider a particular
ground state, denoted by $|0;\Gamma\rangle$, and calculate the
expectation value of $F_p(x)$ in this state.  First we insert the
completeness relation (\ref{completeKleinboson}) between the rightmost
Klein factor and the exponential in Eq. (\ref{Fpboson}), and use
Eq. (\ref{matelneeded}). Upon introducing
$\tilde{\phi}_{+\rho}=\phi_{+\rho}-\langle \phi_{+\rho}\rangle$ etc.,
we encounter the expression
\be
\langle 
\exp{[i(\tilde{\phi}_{+\rho}+s\tilde{\phi}_{+\sigma}-
d_{\lambda\bar{\lambda}}\tilde{\theta}_{-\rho}-sd_{\lambda\bar{\lambda}}
\tilde{\theta}_{-\sigma})]}\rangle.
\label{Bdef}
\ee
By construction, the pinned tilde-fields have zero expectation
values. We also define $\tilde{\theta}_{+\rho}=\theta_{+\rho}$ etc., for
the fields dual to the pinned fields.  As the Hamiltonian is invariant
under a sign change of any of these tilde-fields, and their commutation
relations are invariant under a combined sign change of any two dual
fields, this expectation value is independent of $s$ and
$d_{\lambda\bar{\lambda}}$, as these variables can only be $\pm 1$. A 
qualitative estimate for this expectation value is calculated in Appendix
\ref{variousexp}. Denoting the expectation value by $B$, we obtain
\begin{eqnarray}
\lefteqn{
\langle F_p(x)\rangle = \frac{B}{2\pi\epsilon}\sum_{\lambda s}
d^{\alpha(p)}_{\lambda\bar{\lambda}}
\langle \kappa_{\lambda s}\kappa_{\bar{\lambda}s}\rangle }\nonumber \\
 & & \hspace{-1.1cm}
\exp{[i(\langle \phi_{+\rho}\rangle+s\langle
\phi_{+\sigma}\rangle - d_{\lambda\bar{\lambda}}\langle
\theta_{-\rho}\rangle - s d_{\lambda\bar{\lambda}} \langle
\theta_{-\sigma}\rangle)]}.
\label{expvalFp}
\end{eqnarray}
Here we have suppressed the $\Gamma$ dependence of the expectation
values appearing after the summation sign.  This expression can now be
evaluated for the ground-state configurations for the SF phase in
Table \ref{pinnedvalues} by inserting values for the pinned fields and
using Eqs. (\ref{anticommKlein}) and (\ref{rel1}). The results are listed
in Table \ref{OPresults}. In the SF phase $\langle F_{-1}(x)\rangle$
is nonzero and {\em imaginary}, which implies that the currents are
nonzero. Explicitly, we find
\be
\langle j_{\perp}(m)\rangle = 2(-1)^{\ell}\langle j_{\ell}(m)\rangle =
\mp \frac{B}{2\pi\epsilon} \cdot 8t_{\perp}a^2 (-1)^m,
\ee
where the upper (lower) sign refers to ground state SF 1
(SF 2). Furthermore, $\langle F_1(x)\rangle$ vanishes identically, so
that there is no modulation in neither $\langle k_{\ell}(m)\rangle$
nor $\langle n_{\ell}(m)\rangle $. We have also shown that the first
order contribution to $\langle k_{\ell}(m)\rangle$ is zero in the SF
phase.\cite{unpublished}

\begin{table}
\begin{ruledtabular}
\begin{tabular}{|l|r|r|r|}
Ground state & $\Gamma$ & $\tilde{F}_1$ &  $\tilde{F}_{-1}$ \\ \hline
SF 1  & $\pm 1$ & $ 0$ & $ 4i$ \\
SF 2  & $\pm 1$ & $ 0$ & $-4i$ \\\hline
CDW 1 & $\pm 1$ & $ 4$ & $  0$ \\
CDW 2 & $\pm 1$ & $-4$ & $  0$ \\ 
\end{tabular}
\end{ruledtabular}
\caption{\label{OPresults}
The quantity $\tilde{F}_p \equiv 2\pi\epsilon \langle
F_p(x)\rangle/B$, as calculated from Eq. (\ref{expvalFp}), for the ground
states in Table \ref{pinnedvalues}. The physical properties of these
states are seen to be independent of the ``gauge'' $\Gamma$, as they
should be.}
\end{table}

Finally, we note that the ground-state degeneracy can be broken in a
formal way by adding to the Hamiltonian a term proportional to the
order parameter. Thus, for the SF phase, one can let $H\to H-\eta
j_{\perp}(1)$, where $\eta$ is an infinitesimal constant. Depending on
whether $\eta\gtrless 0$, ground state SF 1 or SF 2 will have the lower
energy. The small imaginary-valued symmetry breaking term perturbs the
purely real-valued Hamiltonian, selecting a particular ground state
which is intrinsically complex-valued with large imaginary components
in the many-body amplitude.

\section{Discussion}
\label{disc}

As an additional check of our calculations, we have also reproduced
the results for the CDW phase found in Ref.~\onlinecite{LBF98}. In
this phase, the signs of $b^{\rho}_{\text{eo}}$,
$b^{\sigma}_{\text{eo}}$, and $u^{\rho}_{\text{ee}}$ are opposite to
the ones given in Eq. (\ref{ratios}). The same bosonic fields are pinned
as in the SF phase, but their expectation values are different. The
CDW phase also has a two-fold degenerate ground state; the
pinned-field configurations we have used are listed in Table
\ref{pinnedvalues}. The rest of the calculation is identical to the
one presented in Sec.~\ref{sf}, including the calculation of $B$ in
Appendix \ref{variousexp}. Our results for $\langle F_p(x)\rangle$ for
the CDW phase are summarized in Table \ref{OPresults}. We find that
$\langle F_1(x)\rangle$ is nonzero and {\em real}, so that $\langle
n_{\ell}(m)\rangle$ is modulated. This phase has no currents, since
$\langle F_{-1}(x)\rangle=0$.

It is perhaps worth commenting more explicitly on how the Klein
factors affect the calculation of the expectation values of the
various order parameters considered in Sec.~\ref{OP}. Two aspects are
important.  First, the $\langle \kappa_{\lambda
  s}\kappa_{\bar{\lambda} s}\rangle$ in Eq. (\ref{expvalFp}) contribute
relative signs to the various terms in the ($\lambda, s$) summation.
These signs are crucial for determining whether $\langle
F_p(x)\rangle$ is nonzero, or if it instead vanishes identically due
to cancellations. Second, if $\langle F_p(x)\rangle$ is nonzero, the
fact that $\langle \kappa_{\lambda s} \kappa_{\bar{\lambda}s}\rangle$
is purely imaginary affects whether $\langle F_p(x)\rangle$ is real or
imaginary, which in turn determines whether the expectation value of a
given order parameter that depends on $\langle F_p(x)\rangle$ will be
nonzero; see Eqs. (\ref{expjperp})-(\ref{expnell}). 

In this paper we have used the so-called ``field-theoretic''
bosonization.\cite{shankar} We have also performed the calculations
using the more rigorous ``constructive'' bosonization\cite{vondelft}
(however, we still neglect the number-changing property of the Klein
factors).  In the latter approach, Eq. (\ref{commphitheta}) is replaced by
$[\phi_{\lambda s}(x),\theta_{\lambda' s'}(x')]=i(\pi/2)
\delta_{\lambda \lambda'}\delta_{ss'}\text{sgn}(x-x')$. Consequently,
the anticommutation between right- and left-moving fermions with the
same band and spin indices must now be taken care of by the Klein
factors, which therefore acquire an additional $R/L$ index. As a
result, 12 different products $\hat{\Gamma}_i$ of four Klein
factors appear in the Hamiltonian. One must identify all relations
between the $\hat{\Gamma}_i$, as these relations put restrictions on
the permissible sets of eigenvalues $\Gamma_i$.\cite{revschulz98} Thus
the treatment of Klein factors is more complicated than in the
field-theoretic approach, where a single operator $\hat{\Gamma}$
appears. However, the final results for the expectation values of the
order parameters are found to be the same.\cite{unpublished}

Our results imply that the SF phase occurs in the phase diagram of a
weakly interacting general SO(5) invariant model on the half-filled
two-leg ladder.\cite{LBF98} However, the basin of attraction of the SF
phase is not restricted to have SO(5) symmetry. In fact, for all bare
couplings studied in Ref.~\onlinecite{LBF98}, including attractive
interactions that {\em break} SO(5) symmetry, it was found that the RG
flow goes to the SO(5) subspace, where the SF phase is one of the
``attracting directions.'' It would be very interesting to undertake a
complete exploration of the parameter space, to see if the SF phase
could possibly be reached from purely repulsive off-site
density-density interactions, supplemented by various spin-exchange
interactions.

Next, we discuss the possibility of SF order away from half-filling.
For generic incommensurate fillings, Umklapp interactions are absent.
Thus the total charge mode $\phi_{+\rho}$ will be gapless (making the
system metallic), so that $\langle \exp{(i\phi_{+\rho})}\rangle$ will
vanish. The currents will then only show
quasi-LRO.\cite{schulz96,orignac,SWA} Strictly speaking, the system is
then no longer in the SF phase, but shows a dominant tendency toward
SF ordering. On the other hand, for commensurate fillings,
higher-order Umklapp interactions are
present,\cite{schulz94,giamarchi97} so that if these interactions are
not irrelevant,\cite{pin} $\phi_{+\rho}$ may be pinned, making true
LRO possible. These conclusions are consistent with those obtained
from symmetry arguments: In the absence of Umklapp interactions, the
Hamiltonian is invariant under the continuous symmetry
$\phi_{+\rho}\to \phi_{+\rho}+c$ (i.e., the constant $c$ can take
arbitrary values), and pinning of $\phi_{+\rho}$ is forbidden by the
Mermin-Wagner theorem.  This theorem no longer applies when Umklapp
interactions are present, since then the symmetry is reduced to a
discrete one (i.e., $c$ can only take particular values).

Finally, we show that our results for the half-filled SF phase are
consistent with the results obtained for the doped ladder for generic
incommensurate fillings. In Table II of Ref.~\onlinecite{orignac} the
values of the three pinned fields in the phase with dominant tendency
toward SF ordering are taken to be $\langle
\phi_{+\sigma}\rangle=\pi/2$, $\langle \theta_{-\rho}\rangle=0$,
$\langle \theta_{-\sigma}\rangle =0$. Using the SF couplings in Eq.
(\ref{ratios}), and taking $\Gamma=-1$, one sees that these expectation
values of the pinned fields indeed minimize $H_{I}^{(1b)}$, and also
$H_{I}^{(2)}$ at half-filling when $\langle \phi_{+\rho}\rangle$ is
taken to be an odd multiple of $\pi/2$.

\begin{acknowledgments}
We are grateful to Ian Affleck, Miguel Cazalilla, Chung-Hou Chung,
Matthew Fisher, Thierry Giamarchi, Tony Houghton, Steve Kivelson,
Pakwo Leung, Hsiu-Hau Lin, Doug Scalapino, Shan-Wen Tsai, Chandra
Varma, and especially Leon Balents and Asle Sudb{\o} for useful
comments and discussions. We also thank Asle Sudb{\o} for interesting
us in the two-leg ladder system, and Leon Balents for telling us about
the bosonization approach to this problem.  J.O.F. was supported by
the Norwegian Research Council, Grant No. 142915/432, and by the
American-Scandinavian Foundation.  J.B.M. was supported in part by the
U.S. National Science Foundation, Grant No. DMR-9712391.
\end{acknowledgments}

\appendix

\section{Matrix elements of products of Klein factors}
\label{matelKlein}

The Hermitian operator $\hat{\Gamma}=
\kappa_{\text{e}\up}\kappa_{\text{e}\down}\kappa_{\text{o}\up}
\kappa_{\text{o}\down}$
enters into bosonic expressions for ${\cal H}_I^{(1b)}$ and ${\cal
H}_I^{(2)}$ [Eqs. (\ref{HI1b})-(\ref{HI2})]. Its eigenvalues are
$\Gamma=\pm 1$, and the associated eigenstates $|\Gamma\rangle$ obey
$\langle \Gamma|\Gamma'\rangle=\delta_{\Gamma\Gamma'}$. The
completeness relation in the space spanned by these eigenstates is 
\be
\sum_{\Gamma=\pm 1}|\Gamma\rangle\langle\Gamma|=I.
\label{completeKlein}
\ee
We want to calculate various matrix elements of 
$\kappa_{\text{e}s}\kappa_{\text{o}s}$, which appear in the 
expectation values of the order parameters considered in Sec.~
\ref{OP}. We have 
\be
\kappa_{\text{e}\up}\kappa_{\text{o}\up}\hat{\Gamma}=
\kappa_{\text{e}\up}\kappa_{\text{o}\up}
\kappa_{\text{e}\up}\kappa_{\text{e}\down}
\kappa_{\text{o}\up}\kappa_{\text{o}\down}
=\kappa_{\text{e}\down}\kappa_{\text{o}\down}.
\ee
Using $\hat{\Gamma}|\Gamma\rangle=\Gamma|\Gamma\rangle$, one obtains 
\begin{eqnarray}
\langle \Gamma|\kappa_{\text{e}\up}\kappa_{\text{o}\up}|\Gamma\rangle &=& 
\Gamma
\langle \Gamma|\kappa_{\text{e}\down}\kappa_{\text{o}\down}|\Gamma\rangle,
\label{rel1}\\
\langle -\Gamma|\kappa_{\text{e}\up}\kappa_{\text{o}\up}|\Gamma\rangle &=& 
\Gamma \langle 
-\Gamma|\kappa_{\text{e}\down}\kappa_{\text{o}\down}|\Gamma\rangle.
\label{rel2}
\end{eqnarray}
The complex conjugate of Eq. (\ref{rel2}) can be rewritten as
$\langle \Gamma|\kappa_{\text{e}\up}\kappa_{\text{o}\up}|-\Gamma\rangle =
\Gamma\langle \Gamma|\kappa_{\text{e}\down}\kappa_{\text{o}\down}|-
\Gamma\rangle$.
 Letting $\Gamma\to -\Gamma$ in this relation, and comparing with
Eq. (\ref{rel2}), shows that the off-diagonal matrix elements are zero;
\be
\langle -\Gamma|\kappa_{\text{e}s}\kappa_{\text{o}s}|\Gamma\rangle=0.
\label{offdiag}
\ee
Next, consider the equation 
$\langle \Gamma|\hat{\Gamma}|\Gamma\rangle = \Gamma$. Anticommuting the
two inner Klein factors and inserting Eq. (\ref{completeKlein}) gives
\be
\sum_{\Gamma'=\pm
1}\langle\Gamma|\kappa_{\text{e}\up}\kappa_{\text{o}\up}|\Gamma'\rangle\langle
\Gamma'|\kappa_{\text{e}\down}\kappa_{\text{o}\down}|\Gamma\rangle = - \Gamma.
\ee
Using Eqs. (\ref{offdiag}) and (\ref{rel1}), we obtain 
$\langle\Gamma|\kappa_{\text{e}s} \kappa_{\text{o}s}|\Gamma\rangle^2=-1$, i.e.
\be
\langle\Gamma|\kappa_{\text{e}s}\kappa_{\text{o}s}|\Gamma\rangle=\pm i.
\label{diag}
\ee
This result is consistent with $\kappa_{\text{e}s}\kappa_{\text{o}s}$ being an
anti-Hermitian operator, thus having a purely imaginary
expectation value.

In order to determine the matrix elements, one can, e.g., fix the two
diagonal matrix elements for one of the spin directions. The other
matrix elements are then determined from Eqs. (\ref{rel1}) and
(\ref{offdiag}). In this paper we choose to set $\langle
\Gamma|\kappa_{\text{e}\up}\kappa_{\text{o}\up}|\Gamma\rangle = +i$.

Finally, we consider the space spanned by the states $|n;\Gamma\rangle$ 
defined in Sec.~\ref{OP}. The completeness relation in this space reads
\be
\sum_{n}\sum_{\Gamma=\pm 1}|n;\Gamma\rangle\langle n;\Gamma|=I.
\label{completeKleinboson}
\ee
We will also need the matrix element 
\begin{eqnarray}
\langle 0;\Gamma|\kappa_{\lambda
s}\kappa_{\bar{\lambda}s}|n;\Gamma'\rangle &=& 
\langle\Gamma|\kappa_{\lambda s}\kappa_{\bar{\lambda}s}|\Gamma'\rangle 
\langle 0(\Gamma)|n(\Gamma')\rangle \nonumber \\ 
 &=& \langle \Gamma|\kappa_{\lambda
s}\kappa_{\bar{\lambda}s}|\Gamma\rangle
\delta_{\Gamma\Gamma'}\delta_{n0}.
\label{matelneeded}
\end{eqnarray}

\section{Expectation values of exponentials of pinned fields}
\label{variousexp}

In this appendix, we calculate a qualitative estimate of the
expectation value in Eq. (\ref{Bdef}) by employing a simplified treatment
of the Hamiltonian, in which the pinned fields are expanded to
quadratic order around their expectation values. This gives
$\langle\Gamma|H|\Gamma\rangle= \sum_{r\nu} H_{r\nu}$ with
$[H_{r\nu},H_{r'\nu'}]=0$, where $H_{r\nu}$ is of Klein-Gordon form,
\begin{eqnarray}
\hspace{-0.7cm}{\cal H}_{+\nu} &=& \frac{1}{2\pi}\Bigl[uK (\partial_x 
\tilde{\theta}_{+\nu})^2 + \frac{u}{K} (\partial_x \tilde{\phi}_{+\nu})^2 
+ w^2\tilde{\phi}_{+\nu}^2\Bigr], 
\label{H+nu}\\
\hspace{-0.7cm}{\cal H}_{-\nu} &=& \frac{1}{2\pi}\Bigl[uK (\partial_x 
\tilde{\phi}_{-\nu})^2 + \frac{u}{K}(\partial_x \tilde{\theta}_{-\nu})^2 
+ w^2\tilde{\theta}_{-\nu}^2\Bigr],
\label{H-nu}
\end{eqnarray} 
where $w^2=24g/(\pi\epsilon^2)$ and 
\be
u=\sqrt{v_F^2-(g/\pi)^2},\qquad K=\sqrt{\frac{v_F-g/\pi}{v_F+g/\pi}}.
\ee
Furthermore, we have defined $\tilde{\phi}_{+\rho}=\phi_{+\rho}-
\langle \phi_{+\rho}\rangle$ etc. for the pinned fields, and
$\tilde{\theta}_{+\rho}=\theta_{+\rho}$ etc. for the fields dual to
them. Written in terms of these fields, $H_{r\nu}$ is independent of
$\Gamma$ (and also of whether we consider the SF or CDW phase). 
Clearly, ${\cal H}_{-\nu}$ can be obtained
from ${\cal H}_{+\nu}$ by letting $\tilde{\theta}_{+\nu}\to
\tilde{\phi}_{-\nu}$ and $\tilde{\phi}_{+\nu}\to
\tilde{\theta}_{-\nu}$. In addition,
\be
[\partial_x \tilde{\theta}_{+\nu}(x),\tilde{\phi}_{+\nu}(x')]=
[\partial_x \tilde{\phi}_{-\nu'}(x),\tilde{\theta}_{-\nu'}(x')].
\ee
Thus all Hamiltonians $H_{r\nu}$ are equivalent. It
therefore suffices to consider, e.g., $H_{+\nu}$. We expand the fields as  
\begin{eqnarray}
\tilde{\phi}_{+\nu}(x,t) &=& \sqrt{uK}\sqrt{\frac{\pi}{L}} 
\sum_{q\neq 0} e^{-\epsilon|q|/2} \frac{1}{\sqrt{2\omega_q}} \nonumber
\\ & & \hspace{-1.0cm}
\Bigl\{a_{+\nu q}e^{i(qx-\omega_q t)}+a^{\dagger}_{+\nu q} 
e^{-i(qx-\omega_q t)}\Bigr\},\\
\partial_x \tilde{\theta}_{+\nu}(x,t) &=&
\frac{i}{\sqrt{uK}}\sqrt{\frac{\pi}{L}} \sum_{q\neq 0}
e^{-\epsilon|q|/2}\sqrt{\frac{\omega_q}{2}} \nonumber \\ 
 & & \hspace{-1.0cm}\Bigl\{a_{+\nu q}e^{i(qx-\omega_q t)}-a^{\dagger}_{+\nu q} 
e^{-i(qx-\omega_q t)}\Bigr\},
\end{eqnarray}
where $\omega_q=\omega_{-q}$, and $a_{+\nu q}$ and
$a^{\dagger}_{+\nu q}$ are canonical boson operators satisfying 
$[a_{+\nu q},a^{\dagger}_{+\nu q'}]=\delta_{qq'}$. These expansions
give the correct equal-time commutation relations and equations of
motion. The Hamiltonian can then be written on diagonal form,
\be
H_{+\nu}=\sum_{q\neq 0}e^{-\epsilon|q|}\omega_q 
a^{\dagger}_{+\nu q}a_{+\nu q},
\ee
with $\omega_q=\sqrt{u^2 q^2+uK w^2}$. 

Next we consider the ground-state expectation value of
$\exp{[ic\tilde{\phi}_{+\nu}(x)]}$, where $c$ is an arbitrary
$c$-number. Let
\be
\tilde{\phi}_{+\nu}\equiv
\Phi_{+\nu}+\Phi^{\dagger}_{+\nu},
\ee 
where
$\Phi_{+\nu}$ ($\Phi^{\dagger}_{+\nu}$) contains the
annihilation (creation) part of $\tilde{\phi}_{+\nu}$. The ground
state expectation value can be written
\be 
\langle \exp{[ic\tilde{\phi}_{+\nu}(x)]}\rangle=\exp{\Bigl(-\frac{c^2}{2}
[\Phi_{+\nu}(x),\Phi^{\dagger}_{+\nu}(x)]\Bigr)},
\label{singleexp}
\ee 
where
\be
[\Phi_{+\nu}(x),\Phi^{\dagger}_{+\nu}(x)]=\frac{\pi}{L}uK\sum_{q>0}
\frac{e^{-\epsilon q}}{\omega_q}.
\ee 
In actuality the coupling constant $g$ is not constant
up to arbitrarily high momenta.  Rather, $g$ is really a
function $g(q)$, with $g(q\to 0)=g$, and $g(q\to\infty)=0$. It follows
that $u$, $K$ and $w$ also become momentum-dependent, and $\omega_q$
acquires an additional momentum dependence. Thus, a more correct 
expression for the commutator is
\be
[\Phi_{+\nu}(x),\Phi^{\dagger}_{+\nu}(x)]=\frac{\pi}{L}\sum_{q>0}
u(q)K(q)\frac{e^{-\epsilon q}}{\omega_q(q)}.  
\ee 
For simplicity, we
will assume that there is a characteristic momentum cutoff $1/\Lambda$
such that for $q\ll 1/\Lambda$, $g(q)$ is well approximated by $g$,
and for $q\gg 1/\Lambda$, $g(q)\approx 0$. Multiplying the integrand by
$[e^{-\Lambda q}+(1-e^{-\Lambda q})]$, we then approximate
$g(q)\approx g$ in the term containing $e^{-\Lambda q}$, and $g(q)\approx
0$ in the term containing $(1-e^{-\Lambda q})$. With $\Lambda\gg
\epsilon$, this gives 
\[
[\Phi_{+\nu}(x),\Phi^{\dagger}_{+\nu}(x)] \approx 
\frac{\pi}{L}\sum_{q>0}\Biggl(uK \frac{e^{-\Lambda q}}{\omega_q} +
\frac{e^{-\epsilon q}-e^{-\Lambda q}}{q}\Biggr)
\]
\be
=\frac{\pi K}{4}[\bm{H}_0(z)-Y_0(z)] + \frac{1}{2}\ln(\Lambda/\epsilon),
\label{commfin}
\ee
where $z\equiv \Lambda w\sqrt{K/u}$. Here $\bm{H}_0(z)$ is a Struve
function, and $Y_{0}(z)$ is a Bessel function of the second kind.

Due to the equivalence of the Hamiltonians $H_{r\nu}$, we have
\be
\langle e^{ic\tilde{\phi}_{+\rho}}\rangle =
\langle e^{ic\tilde{\phi}_{+\sigma}}\rangle =
\langle e^{ic\tilde{\theta}_{-\rho}}\rangle =
\langle e^{ic\tilde{\theta}_{-\sigma}}\rangle.
\label{expsymmetry}
\ee
It then follows from Eq. (\ref{singleexp}) that the expectation value in
Eq. (\ref{Bdef}) is independent of $s$ and $d_{\lambda\bar{\lambda}}$, as
these variables are restricted to be $\pm 1$. Denoting this
expectation value by $B$, we find
\be
B\approx\frac{\epsilon}{\Lambda}\exp{\Bigl\{-\frac{\pi K}{2}
[\bm{H}_0(z)-Y_0(z)]\Bigr\}}.
\label{resB}
\ee
The $\epsilon$ in the prefactor cancels the $1/\epsilon$ in the
prefactor of Eq. (\ref{expvalFp}). In the unpinned limit ($g\to 0$),
$B\approx \sqrt{24g/\pi v_F}\to 0$, while in the limit of maximum
pinning ($g\to \pi v_F$), $B\approx \frac{\epsilon}
{\Lambda}\exp{\left(-\frac{\epsilon}{\Lambda} \sqrt{\frac{\pi
v_F-g}{24g}}\right)} \to \frac{\epsilon}{\Lambda}$.


\begin{thebibliography}{}
\bibitem{pin} Note that since these Umklapp interactions transport
  more than two electrons across the Fermi sea, they are always
  irrelevant with respect to the noninteracting fixed point. However,
  depending on the underlying lattice model and the values of its
  parameters, they may be relevant with respect to the bosonic fixed
  point defined by the quadratic part of the bosonized Hamiltonian
  (including the contribution from the chiral interaction terms).
\end{thebibliography}
\end{document}